\documentclass[twocolumn,secnumarabic,amssymb,nobibnotes,aps,prl,superscriptaddress]{revtex4-1}
\usepackage{amsmath}
\usepackage{amsfonts}
\usepackage{amssymb}
\usepackage{graphicx}
\usepackage{hyperref}
\usepackage{longtable}
\usepackage{dcolumn}
\usepackage{bm}
\usepackage{epstopdf}
\usepackage{wrapfig}
\usepackage{color}

\hypersetup{colorlinks,bookmarksopen,bookmarksnumbered,
citecolor=black,
linkcolor=black,
pdfstartview=green,
urlcolor=black}

\newcommand{\D}{{\rm d}}

\newcommand{\DLD}{D_{\mathrm{LD}}}
\newcommand{\DLO}{D_{\mathrm{LO}}}
\newcommand{\Da}{D_\alpha}

\newcommand{\Lkb}{L_{\bar{k}}}
\newcommand{\Lsigma}{L_\sigma}
\newcommand{\Lres}{L_{\mathrm{res}}}

\newcommand{\etakb}{\eta_{\bar{k}}}
\newcommand{\etasigma}{\eta_{\sigma}}
\newcommand{\etares}{\eta_{\mathrm{res}}}
\newcommand{\etapert}{\eta_{\mathrm{pert}}}
\newcommand{\etanu}{\nu}

\newcommand{\tH}{\hat{t}}
\newcommand{\rH}{\hat{r}}

\newcommand{\Eqref}[1]{Eq.~\eqref{#1}}
\newcommand{\fref}[1]{Fig.~\ref{#1}}

\def\ee{\end{equation}}
\def\be{\begin{equation}}
\def\bmul{\begin{equation}gin{multline}}
\def\emul{\end{multline}}
\def\bea{\begin{eqnarray}}
\def\eea{\end{eqnarray}}

\def\st{{\gamma}}

\begin{document}

\title{Measuring Gaussian rigidity using curved substrates}

\author{Piermarco Fonda}
\affiliation{Theory \& Bio-Systems, Max Planck Institute of Colloids and Interfaces, Am M\"uhlenberg 1, 14476 Potsdam, Germany}
\affiliation{Instituut-Lorentz, Universiteit Leiden, P.O. Box 9506, 2300 RA Leiden, Netherlands}
\author{Sami C.~Al-Izzi}%
\affiliation{School of Physics \& EMBL-Australia node in Single Molecule Science, University of New South Wales, Sydney, Australia}
\affiliation{Department of Mathematics, University of Warwick, Coventry CV4 7AL, UK}
\affiliation{Institut Curie, PSL Research University, CNRS, Physical Chemistry Curie, F-75005, Paris, France}
\author{Luca Giomi}%
\affiliation{Instituut-Lorentz, Universiteit Leiden, P.O. Box 9506, 2300 RA Leiden, Netherlands}
\author{Matthew S.~Turner}%
\affiliation{Department of Physics, University of Warwick, Coventry CV4 7AL, UK}
\affiliation{Centre for Complexity Science, University of Warwick, Coventry CV4 7AL, UK}
\affiliation{Department of Chemical Engineering, University of Kyoto, Kyoto 615-8510, Japan}

\begin{abstract}
The Gaussian (saddle splay) rigidity of fluid membranes controls their equilibrium topology but is notoriously difficult to measure. In lipid mixtures, typical of living cells, linear interfaces separate liquid ordered (LO) from liquid disordered (LD) bilayer phases at subcritical temperatures. Here we consider such membranes supported by curved substrates that thereby control the membrane curvatures. We show how spectral analysis of the fluctuations of the LO-LD interface provides a novel way of measuring the difference in Gaussian rigidity between the two phases. We provide a number of conditions for such interface fluctuations to be both experimentally measurable and sufficiently sensitive to the value of the Gaussian rigidity, whilst remaining in the perturbative regime of our analysis.
\end{abstract}

\maketitle

Bilayer fluid membranes containing mixtures of phospholipid molecules are ubiquitous in cell biology and are used to compartmentalise many of the cell's internal components \cite{alberts_molecular_2002}. Under certain conditions such membranes can phase separate into cholesterol-rich Liquid Ordered (LO) and cholesterol-poor Liquid Disordered (LD) phases, both {\it in vivo} \cite{edidin_state_2003,brown_structure_2000,simons_lipid_2000,simons_cholesterol_2002} and {\it in vitro} \cite{gebhardt_domain_1977,shimokawa2017formation,simons_model_2004}. \textit{In vitro} such a transition is often controlled by decreasing the temperature below some critical value. In the present work we analyse two-phase membranes supported on pre-engineered substrate surfaces, such as have recently been developed \cite{rinaldin2020geometric,rinaldin2019colloid}.

In the sharp interface approximation, the free energy of a multicomponent, demixed fluid membrane can be described by a minimal set of parameters relative to each phase. The in-plane isotropy of fluid membranes demands the free energy to depend only on coordinate-invariant geometric quantities, such as the mean curvature $H$ and the Gaussian curvature $K$ \cite{Canham1970,Helfrich1973,safran2018statistical}.  

For a symmetric bilayer and relatively small curvatures this energy takes the simple form \cite{Julicher1993}
\be
\label{eq:freeEnergy}
F=
\sum_\alpha 
\int_{\Da} \D A \, 
\left(
	\st_\alpha+ 2k_\alpha H^2 + \bar{k}_\alpha K
\right) 
+
\sigma \oint_\Gamma {\rm d}s 
\,,
\ee
here $\Da$ ($\alpha = \mathrm{LO,LD}$) indicates the different phase domains and $\Gamma = \partial \Da$ is the set of closed lines separating them; $dA = dx^1 dx^2 \sqrt{g}$ is the area element, with $\{x^1,x^2\}$ local coordinates and $g$ the determinant of the induced metric $g_{ij} = \partial_i r \cdot \partial_j r$, and $r = r(x^1,x^2)$ the membrane location; $s$ is the arc-length parameter of $\Gamma$ and $\sigma$ is its associated line tension. The mechanical parameters are the surface tensions of each phase $\st_\alpha$, together with their bending $k_\alpha$ and Gaussian $\bar k_\alpha$ rigidities. The contribution from a term involving spontaneous curvature, omitted from \Eqref{eq:freeEnergy}, may safely be neglected without loss of generality for the specific surfaces we consider, see Ref. \cite{SI}.

While there are several techniques to measure the bending rigidity and surface tension of membranes (see e.g. Refs.~\cite{Dimova2014,Nickels2015,Nagle2015}), 
the Gaussian rigidity is much more elusive and very few estimates exist, either from simulations \cite{hu2012determining} or experiment \cite{baumgart2005membrane,semrau2008accurate,rinaldin2020geometric,morris_solvated_2019}. 
The essential difficulty is that the integral $\int K \D A$ is a topological invariant for any closed surface meaning that the Gaussian rigidity does not affect the total free energy of a membrane unless it is subject to topology-changing deformations, such as red fission and fusion.
This topological protection is lost for surfaces with boundaries, since the surface integral can be recast as a line integral of the boundary's geodesic curvature. This case  includes phase-separated multicomponent membranes, where the domains $\Da$ of Eq.~(\ref{eq:freeEnergy}) have a boundary $\Gamma$. The motivation of the present study is to exploit the sensitivity of the linear interface to $K$ in order to measure the Gaussian rigidity difference between the two phases.

While the energy Eq.~(\ref{eq:freeEnergy}) can describe the shape of phase-separated free-standing membranes, such as multi-component vesicles \cite{baumgart2005membrane,milner1987dynamical}, it also applies to the case of curved supported lipid bilayers (SLBs), where the membrane lies on a rigid substrate of non-trivial curvature, while preserving its liquid nature \cite{parthasarathy2006curvature,Subramaniam2010,rozycki2008stable}. These experimental techniques have been recently extended to a wide variety of shapes with both open and closed surfaces in \cite{rinaldin2020geometric,rinaldin2019colloid}. Thus, supporting a membrane using a pre-engineered surface with arbitrary shape is therefore now a realistic proposal \cite{Li2019}. Our goal here is to design substrates that might be used in such a protocol to probe novel physics. We assume that the surfaces are uniformly functionalised to be weakly adhesive.

\begin{figure}[t]
\centering
\includegraphics[width=0.4\textwidth]{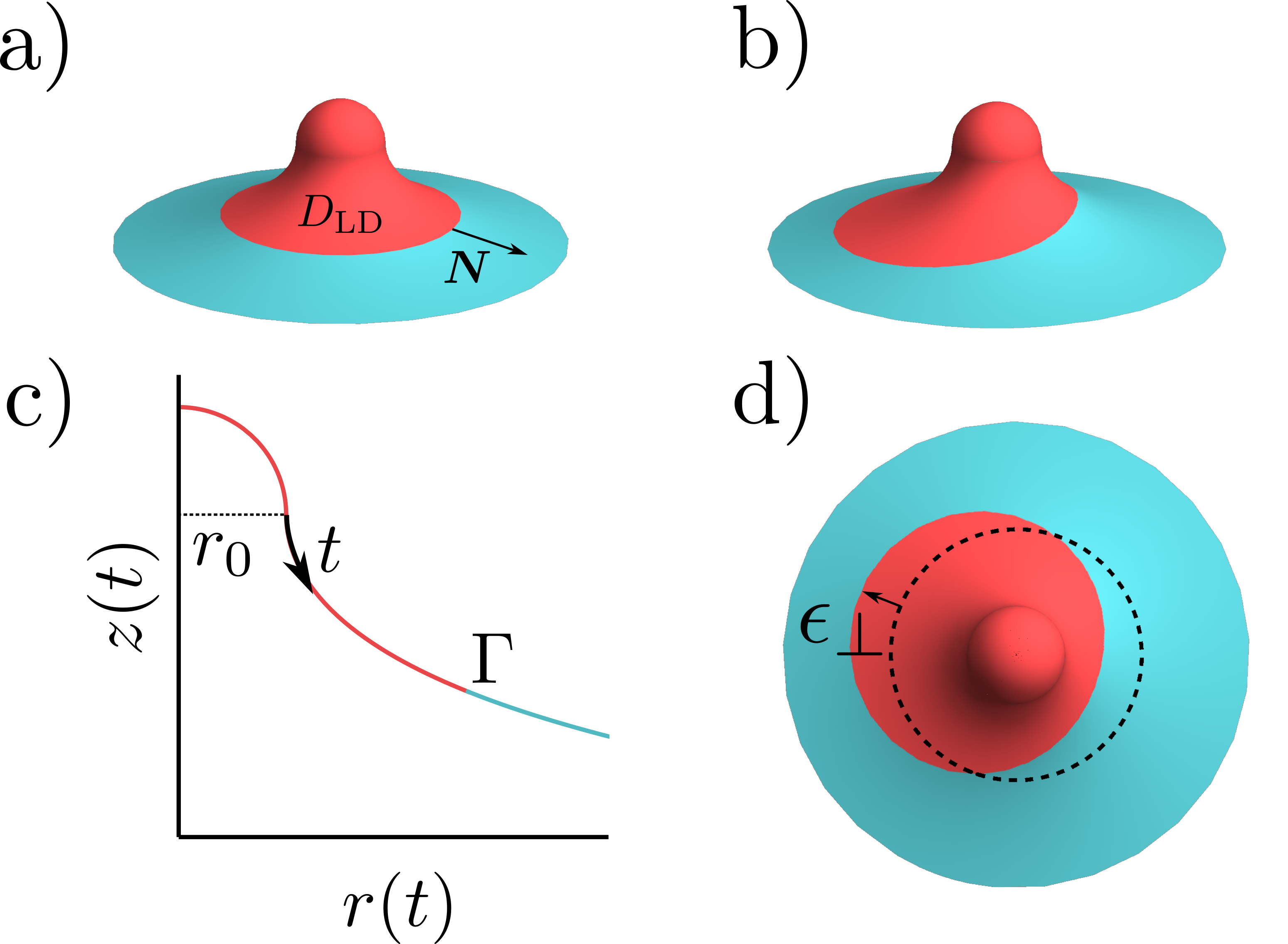}
\caption{A surface with the geometry of a catenoid, closed (arbitrarily) with a spherical cap, serving as a substrate for a bound membrane. A domain of LD (red, labelled $D_{LD}$) is shown surrounded by LO (cyan). (a) The equilibrium LO-LD interface, $\Gamma$, is a symmetric circle. (b) Thermal fluctuations displace the LD domain (the $m=1$ mode shown), the mean squared magnitude of which provides a probe of Gaussian rigidity. (c) The arc length variable $t$, measured from the minimal catenoid radius, $r_0$, along a line of constant $\phi$, parameterises a family of circles of radius $r(t)$ on the surface. (d) The interface displacement projected from above $\epsilon_\perp$, as might be observed experimentally.} 
\label{fig:surfaceSchematic}
\end{figure}

For phase-separated SLBs the only degree of freedom in Eq.~(\ref{eq:freeEnergy}) is the location of the interface $\Gamma$.  If $\Gamma$ is parametrized by a curve $\bm{r}=\bm{r}(s)$ within the surface, where $s$ is the arc-length, one can study the response of the energy to small displacements $\bm{r} \to \bm{r} + \epsilon\>\bm{N}$, with $\bm{N}=\bm{N}(s)$ the so-called tangent-normal vector to $\Gamma$, chosen to point towards $\DLO$ domains (see Fig. \ref{fig:surfaceSchematic}a), and $\epsilon= \epsilon(s)$ a small scalar function with dimensions of a length. 
Since the substrate is assumed smooth and uniform there is no surface discontinuity at the LO/LD interface.
In Ref.~\cite{fonda2018interface} it was shown how to perform a perturbative expansion of the free energy in powers of $\epsilon$ as
\be
F = F_0 + \delta^{(1)}F + \frac{1}{2} \delta^{(2)} F + O(\epsilon^3) \,,
\ee
with 
\be
\label{eq:varF}
\delta^{(1)} F 
= 
\int_\Gamma ds \, \epsilon\left( \sigma \kappa_g -2\Delta k H^2 - \Delta \bar{k} K - \Delta \st \right) \,,
\ee
and
\begin{multline}
\label{eq:secondVarFequilibrium}
\delta^{(2)} F 
= 
\int_\Gamma \D s \Big[ \sigma {\dot\epsilon}^2 - \epsilon^2 \big( \sigma K +\sigma \kappa_g^2
\\+ 2\Delta k \nabla_{\bm{N}} H^2 + \Delta \bar{k} \nabla_{\bm{N}} K \big)\Big] \,,
\end{multline}
where $\Delta k= k_{\rm LO} -k_{\rm LD}$, $\Delta \bar k= \bar k_{\rm LO}-\bar k_{\rm LD}$, $\Delta \gamma= \gamma_{\rm LO}-\gamma_{\rm LD}$ and a dot indicates differentiation with respect to the arc-length: e.g. $\dot{\epsilon}=d\epsilon/ds$. $\kappa_g$ is the geodesic curvature of the interface with the convention that $\kappa_g>0$ for a convex $\DLD$ domain \cite{fonda2018interface}.
Note that \Eqref{eq:secondVarFequilibrium} does not explicitly depend on the tension difference $\Delta \gamma$; in fact, for membranes bound on a fixed support, the only effect of tension is to enforce domains of fixed areas, in exactly the same way that the Laplace pressure enforces a finite volume to a three-dimensional droplet. 

We are interested in thermal fluctuations of $\Gamma$ around a stable equilibrium position, for which $\delta^{(1)} F=0$. Then, from \Eqref{eq:secondVarFequilibrium} it is possible to derive the fluctuation spectrum of the interface. For simplicity we assume a single, simply connected domain $\DLD$ of liquid disordered phase surrounded by a reservoir of LO phase.

We analyse an experimental configuration similar to the one shown in \fref{fig:surfaceSchematic}: an axisymmetric surface and a ground-state interface with circumference $\ell_\Gamma$, so that $\bm{N}$ is along a meridian direction. The rotational symmetry implies that both principal curvatures and their normal derivatives are independent of the arc-length $s$. 
The normal displacement $\epsilon$ along the surface tangent can be decomposed in Fourier modes as
\be
\label{eq:epsilon_def}
\epsilon(s) = \sum_{m=-\infty}^{+\infty} \epsilon_m e^{i \omega_m s},
\ee
with $\epsilon_m$ the amplitude of the mode having discrete wavenumber $\omega_m = 2 \pi m / \ell_\Gamma$. 
Since we are considering the case of fixed $\DLD$ area, the $m=0$ mode is suppressed and $\epsilon_0=0$ \cite{SI}. 

At lowest order the energy is quadratic in the fluctuation amplitudes and we use the equipartition theorem to infer the mean squared amplitude of each mode at temperature $T$:
\be 
\label{eq:generalfluctuations}
\langle|\epsilon_m|^2\rangle
=
\frac{k_B T}{\sigma \frac{8 \pi^2}{\ell_\Gamma} m^2 - 2\ell_\Gamma P  } \,,
\ee
where $P=\sigma(K + \kappa_g^2)+  2\Delta k \nabla_{\bm{N}} H^2 + \Delta \bar{k} \nabla_{\bm{N}} K $ \cite{fonda2018interface} and $k_B$ is Boltzmann's constant. 
This is one of the key results of the present work, since it allows us to link thermally induced displacements of $\Gamma$ to the parameter $\Delta \bar{k}$. Note that the larger the value of $m$, the more suppressed the mode amplitudes, so that only the lowest modes are experimentally accessible. \Eqref{eq:generalfluctuations} generalizes results for planar domains: with a flat surface ($H=K=0$), and $\DLD$ to be a circular domain of radius $r$ so that $\ell_\Gamma=2 \pi r$ and $P= \sigma / r^2$ we recover
\be 
\label{eq:planarfluctuations}
\langle|\epsilon_m|^2\rangle
=
\frac{k_B T}{4 \pi \sigma}
\cdot 
\frac{r}{m^2 - 1} \,,
\ee
in agreement with Ref.~\cite{goldstein1994domain}. 
This relationship has been successfully employed in Refs.~\cite{esposito2007flicker,tian2007line,honerkamp2008line,usery2017line} to measure the line tension $\sigma$ by observing the ``flickering" shape fluctuations of quasi-circular domains on large spherical vesicles. The $m=1$ divergence is due to the fact that $\epsilon_1$ fluctuations are infinitesimal translations, which, being an isometry of the flat plane, cost zero energy. In experiments one removes this mode by using coordinates relative to the domain position. As we show below, this same mode becomes non-trivial on surfaces which do not have translational invariance.

As shown in \fref{fig:surfaceSchematic}, we propose a truncated catenoid as a candidate for a substrate geometry that engineers sensitivity to the Gaussian rigidity. This is primarily because catenoids are minimal surfaces, i.e. with $H=0$, so that \Eqref{eq:freeEnergy} is insensitive to the bending moduli $k_\alpha$ (and to spontaneous curvature, see Ref. \cite{SI}).
We assume that $\DLD$ is a single circular domain centered at the catenoid origin and with the axisymmetric interface lying in the lower branch of the surface. We need a truncated geometry to ensure finiteness of the domain, and we arbitrarily choose to cut away the upper branch of the surface and replace it with a half sphere. The spherical cap should pin $\DLD$ at its desired position because the LD phase is (much) softer.

In arc-length parametric form, the catenoid profile functions can be written as 
\be
\begin{split}
&r(t) = \sqrt{r_0^2+t^2} \,,
\\
&z(t) = z_0 +\frac{r_0 }{2}\log\left(\frac{r(t)-t}{r(t)+t}\right) \,,
\end{split}
\ee
with $r_0$ the neck radius, $z_0$ the neck height and $t$ the interface distance from the catenoid's neck. The geometry is truncated and thus we restrict the possible interface position to the range $t > 0$. The geodesic curvature of $\Gamma$ is constant, $\kappa_g =t/r(t)>0$, hence $\delta^{(1)} F =0$ is satisfied for adequate values of $\Delta \gamma$ \cite{fonda2018interface}. The Gaussian curvature and its normal derivative are respectively
$ K= -r_0^2/r(t)^4$ and $\nabla_{\boldsymbol{N}}K = 4 r_0^2 t/r(t)^6$. The interface length is $\ell_\Gamma = 2 \pi r(t)$ and so $P= \sigma (t^2-r_0^2)/r(t)^4+ 4 \Delta \bar{k} r_0^2 t/r(t)^6$. \Eqref{eq:generalfluctuations} then becomes 
\be 
\label{eq:catenoidfluctuations}
\langle|\epsilon_m|^2\rangle
=
\frac{k_B T}{4 \pi \sigma} 
\cdot 
\frac{r(t)}{m^2 - 1 +2 \frac{r_0^2}{r(t)^2} + 4 \Lkb \frac{r_0^2 t}{r(t)^4}} \,.
\ee
This differs from the planar case. It has two additional terms in the denominator: the first is due to the non-trivial intrinsic geometry of the substrate and the second to the coupling with the Gaussian rigidity, expressed in terms of the length scale $\Lkb = - \Delta \bar{k} / \sigma>0$ (typically of the order of a few hundred nanometers \cite{baumgart2005membrane,semrau2008accurate,rinaldin2020geometric}). Notice that the temperature-dependent prefactor also has dimensions of a length, which we write $\Lsigma = k_B T / 4 \pi \sigma$.

The striking difference between \Eqref{eq:planarfluctuations} and \Eqref{eq:catenoidfluctuations} is that in the latter case the $m=1$ mode has a non-zero energy cost, implying that $\langle|\epsilon_1|^2\rangle$ is a non-diverging quantity. This mode does not correspond to pure translations anymore, but rather involves a ``tilt'' in which the centre of the domain misaligns with the catenoid's axis (see \fref{fig:surfaceSchematic}b and d). In the following, we will focus our discussion only this mode since it has the largest amplitude and should be the most straightforward to extract from data.

In order to be able to measure $\Delta\bar{k}$ experimentally, several criteria must be met, both when designing the geometry of the substrate and when identifying a suitable domain size to measure. We imagine that a future experiment would involve imaging a number of domains pinned to catenoidal substrate features and that these domains would exhibit a broad variation in their area. This means that identifying a domain that is in any appropriate size range should not be a fundamental problem, provided the preferred size is not too restrictively specified.

\begin{figure}
\centering
\includegraphics[width=\columnwidth]{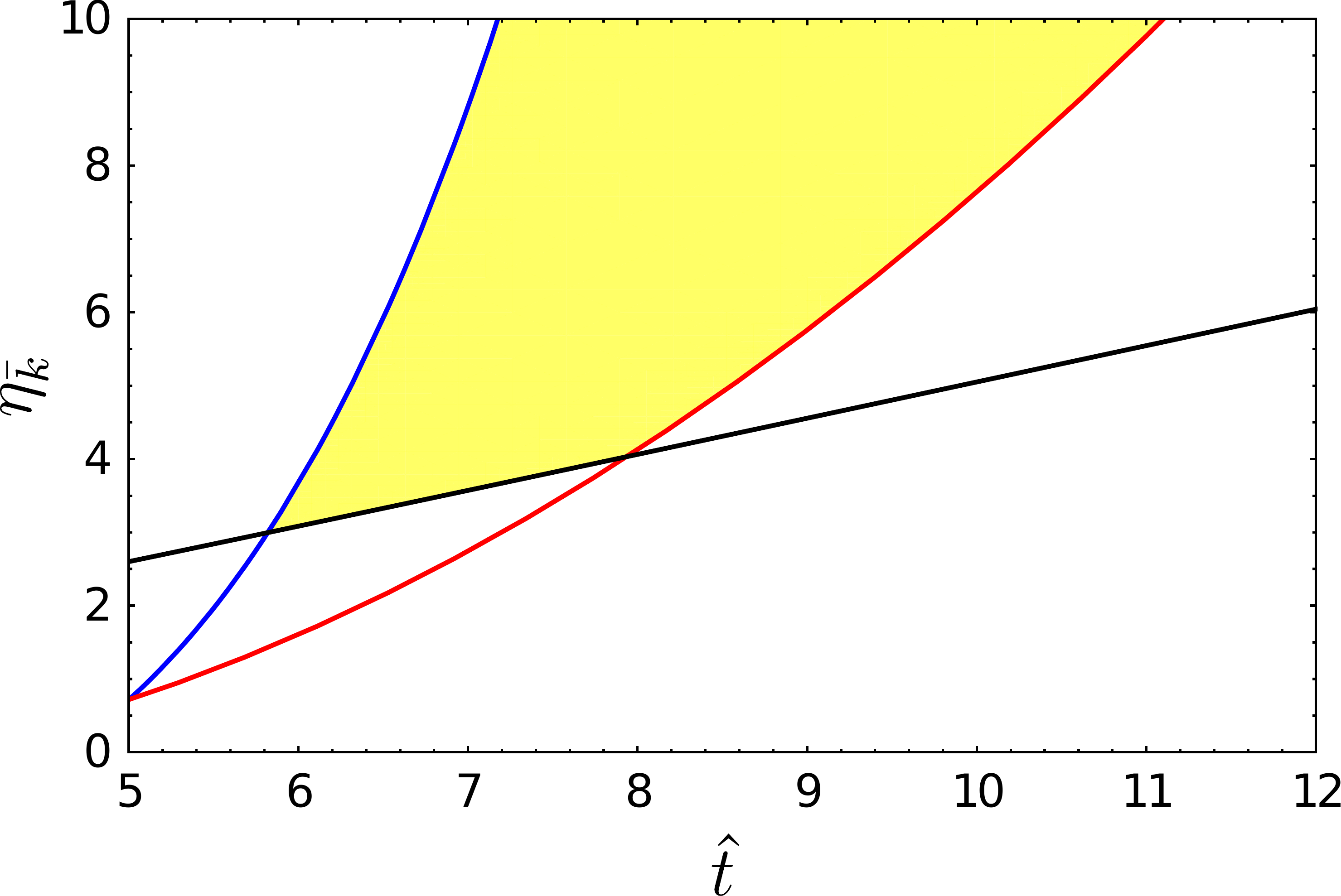}
\caption{Values of the non-dimensionalized coupling $\etakb=- \Delta \bar{k} / (\sigma r_0)$ that can be probed with our method as a function of the rescaled domain size $\tH=t/r_0$, provided all three feasibility constraints are satisfied. In this figure, we take relatively stringent limits on the constraint parameters, as explained in the main text. The blue, red and black line correspond respectively to constraints \textit{(I)}, \textit{(II)} and \textit{(III)}. The yellow region is the intersection of these, and thus shows the values of $\etakb$ that could be measured for a suitable domain size.} 
\label{fig:regionplot1}
\end{figure}

We identify the following three constraints that our system should satisfy:\newline
\textit{(I)} the fluctuations must be experimentally observable, i.e. the amplitude must be larger than a minimal microscope resolution $\Lres$; \newline
\textit{(II)} fluctuations should be small enough so that they can be treated perturbatively - with no significant $O(\epsilon^3)$ effects - and avoid the unpinning of $\DLD$; \newline
\textit{(III)} the $m=1$ mode amplitude should depend sufficiently strongly on $\Lkb$ so that the value of $\Delta \bar{k}$ can then be inferred.

Criterion \textit{(I)} can be enforced by requiring that the fluctuation amplitude has a lower bound $\langle|\epsilon_1|^2\rangle > \Lres^2 r(t)^2/t^2$, where the extra $r(t)/t$ factor takes into account the aberration induced by the projection onto a plane of the curved domain. Criterion \textit{(II)} is enforced by requiring that the tilt should be significantly smaller than the total interface length $\ell_\Gamma=2\pi r(t)$, which we enforce by requiring $\langle|\epsilon_1|^2\rangle \lesssim r(t)^2 \etapert$ with $\etapert$ an arbitrary small parameter. Criterion  \textit{(III)} is enforced by requiring that the fourth term in the denominator of \Eqref{eq:catenoidfluctuations}, depending on $\Lkb$ must not be much smaller than the third one, which does not, i.e. $ 2 \Lkb t/r(t)^2 \gtrsim \etanu$, with $\etanu\sim 1$. 

To make progress, we non-dimensionalize the quantities involved in \Eqref{eq:catenoidfluctuations}. A natural length scale of our problem is the catenoid's neck size $r_0$, so that all other length scales can be measured in terms of this quantity. We thus express the size of $\DLD$ in terms of $\tH \equiv t/r_0$ and $\rH \equiv r(t)/r_0=\sqrt{1+\tH^2}$. Furthermore, we introduce the dimensionless ratios $\etakb \equiv \Lkb/r_0$, $\etasigma \equiv \Lsigma / r_0$ and $\etares \equiv \Lres / r_0$. With this notation, critera \textit{(I)} and \textit{(II)} can be written as
\be 
\frac{2 \etares^2}{\etasigma} \frac{1}{\tH^2} < \frac{\rH^3}{\rH^2+ 2 \etakb \tH} \lesssim \frac{2\etapert}{\etasigma} \,,
\label{eq:condIandII}
\ee
while condition  \textit{(III)} is 
\be 
 \frac{2 \etakb\tH}{\rH^2} \gtrsim \etanu \,.
\label{eq:condIII}
\ee	
If it is possible to satisfy the three inequalities, Eqs. \eqref{eq:condIandII} and \eqref{eq:condIII}, simultaneously, for a given value of $r_0$, $t$ and $\Lkb$, then our approach is experimentally feasible and can be used to measure $\Delta \bar{k}$.

Experiments are assumed to be carried out near room temperature ($T\approx300$K) and the typical values for the line tension span $\sigma \sim 0.1-3.0 \mathrm{pN}$ \cite{usery2017line}. This gives $\Lsigma\sim 1 \mathrm{nm}$. By assuming that the microscope resolution $\Lres$ is about $0.1\mu \rm{m}$ and the catenoid neck $r_0 \sim 0.2 \mu \rm{m}$, we obtain $\etasigma \sim 0.005 $ and $\etares \sim 0.5$. Furthermore, we allow the fluctuation to be roughly $10\%$ of the domain perimeter (i.e. $\etapert = 0.01$) and require a signal to noise ratio of $\etanu \sim 1$.  The result of these assumptions is shown in \fref{fig:regionplot1}: the boundary of each constraint is shown as a line (blue, red, black for \textit{(I)}, \textit{(II)} and \textit{(III)} respectively) and their intersection is highlighted in yellow.

Interestingly, we find that the minimum allowed value for $\etakb$ to be measurable is about $3$, which, with an $r_0$ of $200 \mathrm{nm}$, implies an $\Lkb$ of $600 \mathrm{nm}$, a value well within the range of previous estimates \cite{rinaldin2020geometric}. Furthermore, by relaxing the value of $\etanu$ to $0.5$, this limit is shifted to $200 \mathrm{nm}$, a value below the lowest known estimate \cite{baumgart2005membrane}.

\begin{figure}[t]
\centering
\includegraphics[width=\columnwidth]{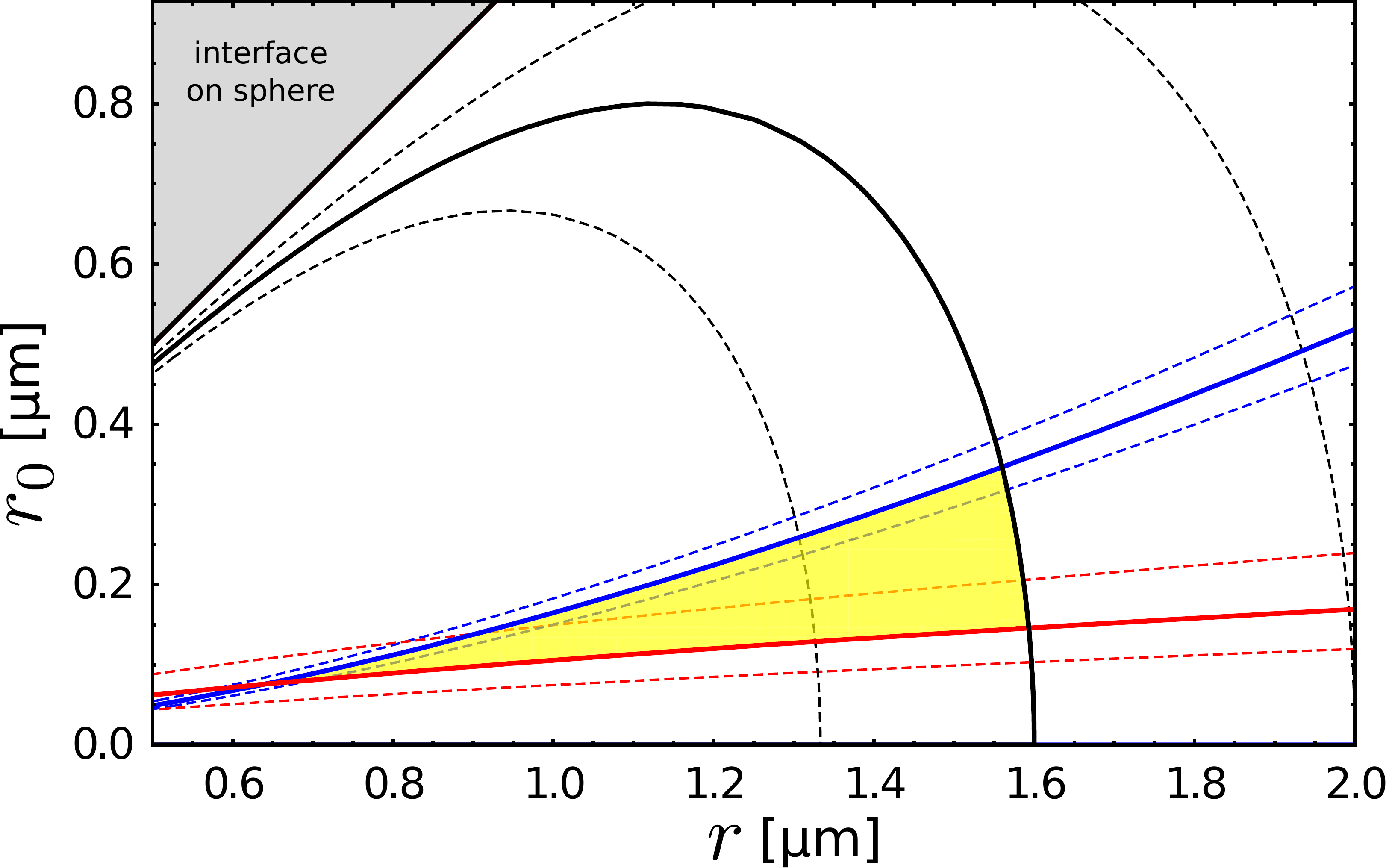}
\caption{The yellow region indicates the space of experimental systems that could allow for accurate measurement of the Gaussian rigidity contrast between two membrane phases by measuring the $m=1$ displacement mode sketched in \fref{fig:surfaceSchematic}, under the assumption $\Lkb= 400 \mathrm{nm}$ and $\Lsigma = 1 \mathrm{nm}$. The catenoidal surface feature(s) have a neck radius $r_0$ and the equilibrium location of the interface is at (projected) distance $r$ from the symmetry axis. The solid lines correspond to criteria \textit{(I)} with $\Lres=100 \mathrm{nm}$ (blue), \textit{(II)} with $\etapert=1/40$ (red) and \textit{(III)} with $\etanu=0.5$ (black). Dashed lines show variations of these values of respectively $\pm 10 \%$, $\pm 20 \%$ and $\pm 50 \%$. We disregard $r<r_0$ as the interface would lie on the spherical cap.
} 
\label{fig:regionplot2}
\end{figure}

While the above discussion and \fref{fig:regionplot1} verify that our proposed system could, in principle, probe the value of $\Delta \bar{k}$, it is also interesting to give an estimate of the geometric features that our system should have in order to probe known values of $\Lkb$. If we take $\Lkb = 0.4 \mathrm{\mu m}$ \cite{rinaldin2020geometric} we can plot a similar diagram to \fref{fig:regionplot1} but now in terms of the dimensional parameters $r_0$, the catenoid size, and $r$, the required domain size. The result is shown in \fref{fig:regionplot2}. This re-casting again shows that our proposed method is feasible and could be implemented with recent sub-micrometer 3D printing methods \cite{Seniutinas2018,Jia2020}. It may also be possible to create an approximation to a catenoid by draping a lipid bilayer over a short pillar with radius $r_0$; the geometry would be approximately catenoidal and our results would be valid up to leading order in the shape.

Ultimately we aim to extract information on the Gaussian rigidities from the fluctuation amplitude. \Eqref{eq:catenoidfluctuations} can in fact be rearranged to provide an explicit expression for the Gaussian rigidities in terms of the $m=1$ fluctuation amplitude as follows
\be
\Delta\bar{k} 
=
\sigma \frac{r(t)^2}{2 t} - \frac{k_B T}{4 \pi} \cdot \frac{r(t)^5}{4 r_0^2 t \langle|\epsilon_1|^2\rangle
}
\,.
\ee
Our method suggests a possible precision of $\pm 10\%$ on the value of the Gaussian rigidity difference, essentially limited by precision on line tension $\sigma$. As our method relies only on the ``translational/tilt'' mode and not shorter wavelength fluctuations it may be possible for the accuracy of this method to be further improved by the use of super-resolution microscopy techniques \cite{thompson_precise_2002,yildiz_fluorescence_2005}, significantly pushing the yellow region of \fref{fig:regionplot2} towards higher values of $r_0$.

In summary, we have presented a method to estimate the Gaussian rigidity difference between two membrane phases, a quantity that is otherwise notoriously difficult to measure. We have shown that, subject to realistic constraints on the resolvability of the fluctuations, it is possible to design a surface that can be used to infer $\Delta\bar{k}$ whilst still remaining in the perturbative regime of our equations. Our method has an inherent advantage over current approaches to measuring the Gaussian rigidity because it is independent of other parameters, such as bending rigidity and spontaneous curvature. The access to Gaussian rigidity that this assay provides could advance the study of its role in complex domain formation processes \textit{in vivo}.

\acknowledgments{The work of LG, PF was supported by the VIDI grant scheme of
the Netherlands Organisation for Scientific Research (NWO/OCW). PF would like to thank the IAS and the University of Warwick for their support and hospitality during the completion of this project. SCAI~acknowledges funding from the UK EPSRC under grant number EP/L015374/1 (Centre for Doctoral Training in Mathematics for Real-World Systems) and support from the Labex CelTisPhyBio (ANR-11-LABX-0038, ANR-10-IDEX-0001-02). MST Acknowledges the generous support of the JSPS, via a long term fellowship, and the peerless hospitality of Prof Yamamoto at Kyoto University where this work was completed. Finally, all authors would like to extend their sincere thanks to Daniela Kraft and Melissa Rinaldin for discussions and access to unpublished data.}

\bibliography{references}

\end{document}